\documentclass[prl,floatfix,superscriptaddress,twocolumn]{revtex4}
\usepackage{latexsym}
\usepackage{amstext,amsfonts}
\usepackage{epsfig}

\begin{document}

\title{Structural crossover of polymers in disordered media}

\author{Roni Parshani}
\affiliation{Minerva Center \& Department of Physics, Bar-Ilan University, Ramat Gan, Israel}
\author{Lidia A. Braunstein}
\affiliation{Instituto de Investigaciones F\'isicas de Mar del Plata (IFIMAR)-Departamento de F\'isica, Facultad de Ciencias Exactas y Naturales, Universidad Nacional de Mar del Plata-CONICET, Funes 3350, (7600) Mar del Plata, Argentina.}
\affiliation{Minerva Center \& Department of Physics, Bar-Ilan University, Ramat Gan, Israel}
\author{Shlomo Havlin}
\affiliation{Minerva Center \& Department of Physics, Bar-Ilan University, Ramat Gan, Israel}

\date{\today}

\begin{abstract}
We present a unified scaling theory for the structural behavior of polymers embedded in a disordered energy substrate. An optimal polymer configuration is defined as the polymer configuration that minimizes the sum of interacting energies between the monomers and the substrate. The fractal dimension of the optimal polymer in the limit of strong disorder (SD) was found earlier to be larger than the fractal dimension in weak disorder (WD). We introduce a scaling theory for the crossover between the WD and SD limits. For polymers of various sizes in the same disordered substrate we show that polymers with a small number of monomers,
$N \ll N^*$, will behave as in SD, while large polymers with length $N \gg N^*$ will behave as in WD.
This implies that small polymers will be relatively more compact compared to large polymers even in the same substrate.
The crossover length $N^*$ is a function of $\nu$ and $a$, where $\nu$ is the percolation correlation length exponent and $a$ is the parameter which controls the broadness of the disorder.
Furthermore, our results show that the crossover between the strong and weak disorder limits can be seen even within the same polymer configuration. If one focuses on a segment of size $n \ll N^*$ within a long polymer ($N \gg N^*$) that segment will have a higher fractal dimension compared to a segment of size $n \gg N^*$.
\end{abstract}
\maketitle

The study of polymers \cite{PolymerReptation,polymerDynamics,DNA,ProteinFolding,PorousMedia,SpinGlasses,Chakrabarti,Smailer1993} in the presence of disordered media is of broad scientific interest and is relevant to many fields such as protein folding \cite{ProteinFolding}, polymers in porous media
\cite{PorousMedia} and spin glasses \cite{SpinGlasses}. It also has many important applications such as in enhanced oil recovery,
drug delivery and DNA sorting \cite{DNA}. A polymer embedded in a disordered energy substrate, at low temperatures, will settle
down in the optimal configuration, i.e. the configuration with the minimum energy. 

In this paper we study the scaling of the structural properties of such an optimal configuration.
A linear polymer of $N$ monomers, under disorder, can be modeled by an $N$ step self avoiding walk (SAW) on a lattice where each site (or bond)
in the lattice is assigned an energy $\epsilon$, taken from a given distribution. The optimal configuration in such a model is the SAW of length $N$ for which the sum of the energies along its path is minimal.
For generating a broad disorder it is common to use the distribution $P(\epsilon )= 1/(a \epsilon)$ ($\epsilon < e^a$) where the parameter $a$ controls the broadness of the disorder
 \cite{cieplak,Porto,DisorderBehvior2,Braunstein2002}.
Such a distribution generates an exponential disorder where the energy value of each site $i$ on the lattice is given by  $\epsilon_i=\exp(ar_i)$
 where $r_i$ is a random number taken from a uniform distribution between $[0,1]$.
For $a \rightarrow \infty$ we obtain the strong disorder (SD) limit and for small values of $a$ the weak disorder (WD) limit.
In the WD limit essentially all sites contribute to the total sum of $\epsilon_i$, while in the SD limit the total
sum is dominated by a single site with the maximum energy \cite{cieplak,barabasi}. \\
Smailer {\it et. al.} \cite{Smailer1993} studied the properties of a linear polymer in the WD limit using both a uniform and a Gaussian energy distribution. They found that for 2D the end-to-end distance of the polymer, $R$, scales with $N$ as $N \sim  R^{d_{opt}'}$,
where the fractal dimension is $d_{opt}' \cong  1.25$. This result is different from that found by Braunstein
{\it et. al.} \cite{Braunstein2002} for the SD limit. Braunstein {\it et. al.} used the exponential disorder ($\epsilon_i=\exp(ar_i)$) and obtained $N \sim  R^{d_{opt}}$ with $d_{opt} \cong  1.5$. Thus, the polymer is more compact in the presence of strong disorder compared to weak disorder.

In this paper we present a unified scaling theory for the crossover between WD and SD limit.
We claim that the crossover between strong and weak disorder for the optimal polymer problem depends on the characteristic size $a^\nu$
where $\nu$ is the correlation length exponent from percolation theory \cite{Percolation,Percolation2,Percolation3}. 

This claim can be explained as follows:
Consider an infinite lattice where each site $i$ has a value $p_i$ where $p_i\in[0,1]$. In a percolation process an increasing fraction $p$ of the sites with the lowest values, are occupied (the others are removed) until the point $p=p_c$ is reached where the lattice undergoes a phase transition. For $p > p_c$  an infinite number of the lattice sites are connected while for $p < p_c$ the lattice is separated into small finite clusters.

Now assume each site is associated with an energy $\epsilon_i = exp(ap_i)$ and an infinite polymer is observed on the lattice. In SD the highest energy, $\epsilon_{max}$, along a polymer configuration dominates the total energy. The energy $\tilde{\epsilon}_{max}$ which is the highest energy along the {\it optimal} polymer configuration, cannot be smaller than $exp(ap_c)$ and cannot be larger than $exp(ap_c)$ and therefore must be equal to $exp(ap_c)$. If $\tilde{\epsilon}_{max} < exp(ap_c)$ then according to percolation theory only a cluster with a finite number of sites exists and an infinite polymer cannot be embedded inside this cluster. If  $\tilde{\epsilon}_{max} > exp(ap_c)$ then according to percolation theory the polymer is not optimal since there exists a polymer with a lower energy configuration for which $\tilde{\epsilon}_{max} \equiv  exp(ap_c)$. Thus, the optimal configuration is achieved for $\tilde{\epsilon}_{max} \equiv  exp(ap_c)$ since $p_c$ is the lowest value for which an infinite cluster that all its sites have energies lower than $\tilde{\epsilon}_{max}$ exists.

For a finite lattice of size $R$, the value $p_c$ for which the percolation transition occurs is distributed as a narrow distribution with a mean $p_c$ and a standard deviation of $\sigma = {R}^{-1/\nu}$ \cite{OptimalPath}. Let $\epsilon _1 = exp(ap_1)$ be the largest energy and $\epsilon _2 = exp(ap_2)$ the second largest energy on the lattice.
Since $log(\frac{\epsilon_1}{\epsilon_2}) \sim a(p_1-p_2)$ and $p_1$ and $p_2$ are both from the same narrow distribution it follows that $log(\frac{\epsilon_1}{\epsilon_2}) \sim a(p_1-p_2) \sim a\sigma \sim a{R}^{-1/\nu}$.
Therefore, the SD limit ($\epsilon_1 / \epsilon_2 \gg 1$) is obtained for $R \ll a^\nu$ while the WD limit is obtained for $R \gg a^\nu$.

\begin{figure}
\begin{center}
\epsfig{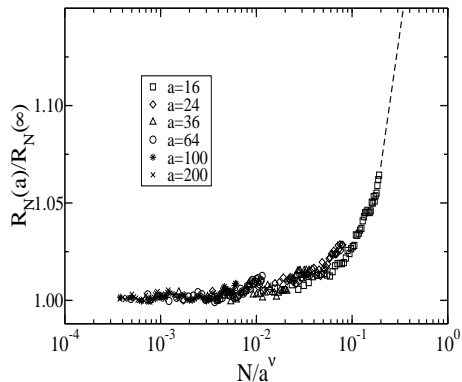}
\end{center}
\caption{Simulation results for the scaling of $R_N(a) / R_N(\infty)$
as function of  $N /a^{\nu'}$ for different values of $a$. The
dashed lines represent the asymptotic expected value of the slope
$x^{1/d_{opt}'-1/d_{opt}}$ (Eq. (\ref{R_N_scaling})).} \label{Fig1}
\end{figure}

Similar considerations have been presented by Buldyrev {\it et. al.} \cite{OptimalPath} for the crossover from SD to WD in the optimal path problem.
Yet, the optimal polymer problem is significantly different from the optimal path problem.
The optimal path problem explores the optimal path between two fixed points (fixed $R$) where the path length varies.
In the optimal polymer problem, the polymer length is fixed and one tries to minimize the sum of the interaction energies between the monomers and the substrate, while the end-to-end distance varies.
While the complexity of the optimal path problem is $O(N^2)$, finding the optimal polymer is an NP (Non-deterministic Polynomial time) problem since nearly all the possible configurations need to be explored in order to find the optimal configuration. To reduce the computational time of our simulations we have used the following optimizations:
a) During the calculation of the optimal polymer of length $N$, the optimal polymers of length $1$ to $N$ are also calculated. b) A new site in the lattice is explored only if the path until that point does not exceed the optimal total energy of length $N$ obtained until that point.
The second optimization is only useful for the SD limit, where one energy dominates the total sum \cite{Note1}. On the other hand for the SD limit
the accuracy of the floating point (double) in conventional computers is limited to $a \sim 36$. We have overcome this limitation by using external software that enables unlimited floating point accuracy, but demands significantly more computational time. These computational problems are the reasons why our simulation results are limited to polymer configurations of length $N=50$.
\begin{figure}[h]
\begin{center}
\epsfig{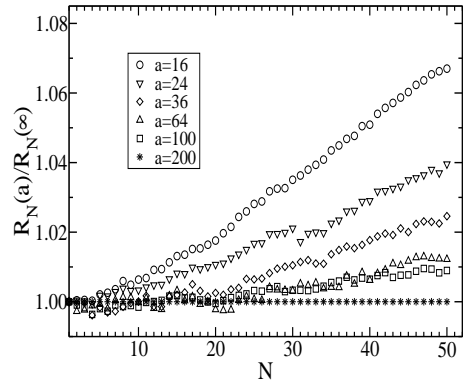}
\end{center}
\caption{Simulation results of $R_N(a)/R_N(\infty)$ as function of
$N$ for several values of $a$.
For small values of $N$ all the polymers behave as in the SD limit and independent of the disorder have the same
fractal dimension. For larger values of $N$, $R_N(a)/R_N(\infty)$ increases towards the WD region.
The results are limited to $N=50$ because calculating $R(a)$ for large $a$ slows down he calculations significantly} \label{Fig3}
\end{figure}

Next, we propose a single scaling function for the dependence of
$R \equiv R_N(a)$ on $N$ that includes both the SD and WD regions. We expect that the scaling function will depend on the ratio $R/a^{\nu}$ which represents the relative strength of the disorder. Since the polymers are of length $N$ we use instead of $R/a^{\nu}$ the scaling parameter 
$N/a^{\nu'} \equiv  N/a^{\nu d_{opt}}$ which is related to $R/a^{\nu}$ by $\frac{R}{a^\nu} \equiv \frac{N^{1/d_{opt}}}{a^{\nu}} \equiv (\frac{N}{a^{\nu d_{opt}}})^{1/d_{opt}}$.
Thus, we propose that
\begin{equation}
R_N(a) \sim R_N(\infty) f\left(\frac{N}{a^{\nu' }}\right) ,
\label{R_N_scaling}
\end{equation}

with $f(x)$ a scaling function given by
\begin{equation}
f(x) = \left\{ \begin{array}{ll} \mbox{const.} & \mbox{$x\ll 1$} \\
x^{1/d_{opt}'-1/d_{opt}} & \mbox{$x\gg 1$} \end{array} \right.
\label{R_N_scaling_function}
\end{equation}

where $d_{opt} \simeq 1.5$ (SD) , $d_{opt}' \simeq 1.25$ (WD) and $R_N(\infty) \sim N^{1/d_{opt}}$.\\
Indeed from Eq. (\ref{R_N_scaling}) and (\ref{R_N_scaling_function}) follows that for $N \ll a^{\nu'}$ (SD) $R_N \sim N^{1/d_{opt}}$, while for $N \gg a^\nu$ (WD) limit we obtain $R_N \sim N^{1/d_{opt}'}$ the known result for the WD limit.
Fig.~\ref{Fig1} presents simulation results supporting the scaling function of Eq. (\ref{R_N_scaling}) and (\ref{R_N_scaling_function}). The
SD region can be seen clearly in the flat area for $N/a^{\nu'} \ll 1$, while the sharp increase in the slope indicates the crossover to WD  until the point where the slope fits the asymptotic expected value, ${1/d_{opt}'-1/d_{opt}}$ (dashed line), see Eq. (\ref{R_N_scaling_function}).

Until now we have been relating to SD and WD as if the polymers
were in different substrates. But actually the crossover between the SD and WD is present
even on the same energy substrate as can be seen from the scaling function of Eq. (\ref{R_N_scaling}) and (\ref{R_N_scaling_function}). In other words, for a given $a$, a small enough polymer ($N \ll a^{\nu'}$) will behave as in SD, 
while a large enough polymer ($N \gg a^{\nu'}$) will behave as in WD. This property is demonstrated in Fig.~\ref{Fig3}
where we plot the ratio $R_N(a)/R_N (\infty)$ as a function of N for different values of $a$.
For small values of N the polymers behave as in SD since $R_N(a)/R_N (\infty)\simeq 1$ (recall that $R_N (\infty)$ is the SD limit). In contrast, for
$N \gg 1$ a crossover towards WD occurs represented by the larger $R$ values. Note, that the crossover value $N^*(a)$ increases for larger $a$.

\begin{figure}[h]
\begin{center}
\epsfig{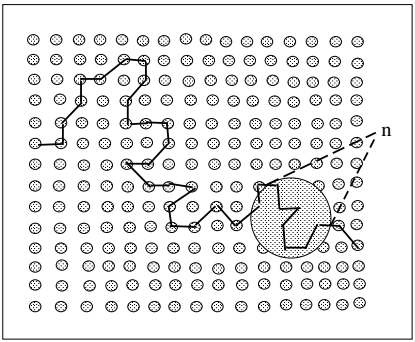}
\end{center}
\caption{ (a) Illustration for the crossover between WD and SD inside the
same polymer. Inside a segment of length $n<a^{\nu'}$ the polymer is more compact (has a higher fractal dimension) compared
to a larger segment or the full polymer.} \label{Fig4}
\end{figure}

\begin{figure}[h]
\begin{center}
\epsfig{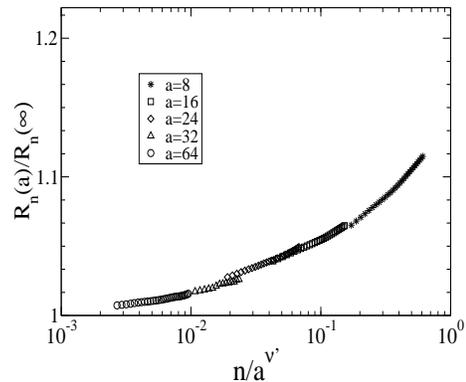}
\end{center}
\caption{ Simulation results for the scaling of $R_n(a) / R_n(\infty)$
as function of  $n /a^{\nu'}$ for different values of $a$. The different values of $n$ represent different segments within the same polymer.
} \label{Fig5}
\end{figure}

 Next we show that such a crossover also exists within the same polymer configuration. We argue that a segment of monomers of length $n \ll a^{\nu'}$, within a polymer of length $N \gg a^{\nu'}$,  will behave as in SD and will be more compact compared to a segment of size ${n'} \gg a^{\nu'}$ that will behave
 as in the WD limit. An illustration of this property is shown in Fig.~\ref{Fig4}.
In Fig.~\ref{Fig5} we present simulation results showing that the same scaling relations describing the crossover between SD and WD for
polymers of different sizes is also correct within the same polymer. Here the simulations were performed for different segments of size $n$ of the same polymer and $R_n(a)$ is the end-to-end distance of these segments.

\begin{figure}
\begin{center}
\epsfig{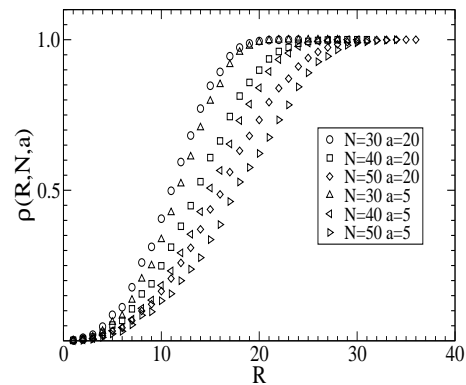}
\end{center}
\caption{Simulation results for the cumulative probability density function, $\rho
(R,N,a)$ as function of
$R$ for different values of $N$ and $a$ in the region $0.075 \leq N/a^{\nu'} \leq  2$. \label{Probability}}
\end{figure}

\begin{figure}
\begin{center}
\epsfig{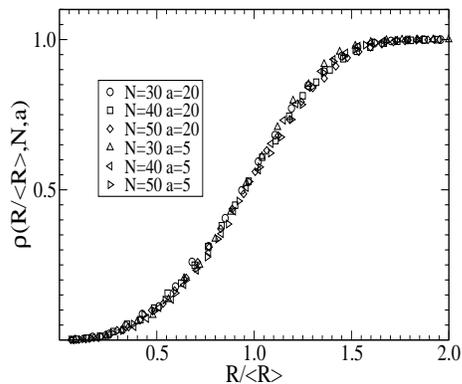}
\end{center}
\caption{Simulation results for the cumulative probability density function, $\rho(R/ \langle R \rangle,N,a)$  as a function of $R/ \langle R \rangle$. When normalizing by $R/ \langle R \rangle$ the different curves collapse into a single curve.\label{ProbabilityScaling}}
\end{figure}

When studying the probability density function of polymers $\rho \equiv \rho (R,N,a)$ we obtain an interesting result:
The width of the distribution does not depend on the crossover parameter of the disorder $N/a^{\nu'}$, only on the normalization parameter $R/\langle R\rangle$.
This is different from the results obtained in optimal path where the probability density function depends on the disorder crossover parameter\cite{OptimalPath}. This difference can be understood by the fact that the polymer configuration is of fixed length $N$. While the length of an optimal path between any two nodes within a distance $R$ may be very long, enabling a broader distribution of the path lengths in SD, the end-to-end distance $R$ in the optimal polymer configuration is limited from above by the fixed length $N$ (in optimal path one measures the distribution of the length, $N$, while in polymers we measure the distribution of $R$). Therefore, the scaling relation for the probability density function of the optimal polymer depends only on the normalization parameter $R/\langle R\rangle$ while the dependence of $N$ and $a$ are determined by $\langle R \rangle$:

\begin{equation}
\rho (R,N,a) = f(\frac{R}{\langle R \rangle >})
\end{equation}

Indeed, Fig.~\ref{Probability} shows simulation results for the cumulative distribution of $\rho (R,N,a)$ for different values of $R,N,a$ that correspond to both the strong and weak disorder regions.
In Fig.~\ref{ProbabilityScaling} the scaling of these lines is presented where all the different distributions both for SD and WD collapse into a
single curve.

Summarizing, we have presented a unified scaling theory for the optimal polymer configuration, embedded in a disordered energy substrate. The structure of the polymer at the minimal energy depends on the strength of the disorder. We find that the crossover between the SD limit and the WD limit occurs at $N^*(a) = N/a^{\nu'}$. 
For $N/a^{\nu'} \gg 1$ the WD is obtained while for $N/a^{\nu'} \ll 1$ the SD limit is obtained.
We present simulation results showing that this transition occurs even inside the same polymer. Therefore, a segment of $n$ monomers where $n/a^{\nu'} \ll 1$ (the SD limit region) will be more compact compared to the full polymer where $N/a^{\nu'} \gg 1$ (the WD region).

Our results of the crossover between the SD limit and the WD limit may be observed in neutron scattering experiments.
Since the structure factor decays with a power (equal to the fractal dimension) of the wave vector, we expect to see a crossover in the structure factor at the wave length for which the transition between the SD and WD occurs. For 3D, the crossover would occur from $d_{opt} \cong  1.82$ in SD limit \cite{Braunstein2002} to $d_{opt}' \cong  1.4$ in the WD limit \cite{Smailer1993}.     
Recently, experimental studies of the scaling properties of the linear chain configuration
of DNA knots adsorbed onto a mica surface under weak trapping \cite{DNAKnots} have shown that $R$ scales with $N$ with a 
fractal dimension of $1.51$. This value is significantly different from the classical Flory result $d_f = 4/3$. The fractal dimension $d_f=1.51$ found in 2D might be explained by Eq. (\ref{R_N_scaling}) in the limit of strong disorder.

We thank the Israel Science Foundation, the Center for Complexity Science (Israel) and LAB thanks UNMdP for financial support.

\end{document}